\RequirePackage{lineno}
\documentclass[twocolumn,showpacs,superscriptaddress,amsmath,amssymb,nofootinbib]{revtex4-2}

\usepackage{graphicx}
\usepackage{subfigure}
\usepackage{subcaption}
\usepackage{epsfig}
\usepackage{overpic}
\usepackage{dcolumn}
\usepackage[normalem]{ulem}
\usepackage{bm}
\usepackage{color}
\usepackage{lineno}
\usepackage{xspace}
\usepackage{epstopdf}
\usepackage{xcolor}
\usepackage{soul}
\usepackage{verbatim}
\usepackage{enumitem}
\usepackage{todonotes}
\usepackage{cancel}
\usepackage{array}
\usepackage[toc,page]{appendix}
\usepackage[colorlinks,allcolors=black]{hyperref}
\usepackage{multirow}
\usepackage{ytableau}
\usepackage{comment}

\usepackage{diagbox}

  \newcommand{\moe}{\affiliation{Key Laboratory of Atomic and Subatomic Structure and Quantum Control (MOE), Guangdong-Hong Kong Joint Laboratory of Quantum Matter, Guangzhou 510006, China
}}

\newcommand{\sfim}{\affiliation{Guangdong Basic Research Center of Excellence for Structure and Fundamental Interactions of Matter, Guangdong Provincial Key Laboratory of Nuclear Science, Guangzhou 510006, China}}

\newcommand{\ihep}{\affiliation{Institute of High Energy Physics, Chinese Academy of Sciences, Beijing 100049, China}}
\newcommand{\ccast}{\affiliation{China Center of Advanced Science and Technology, Beijing 100190, China}}
\newcommand{\iqm}{\affiliation{State Key Laboratory of Nuclear Physics and Technology, Institute of Quantum Matter, South China Normal University, Guangzhou 510006, China}}

\newcommand{\scnt}{\affiliation{Southern Center for Nuclear-Science Theory (SCNT), Institute of Modern Physics, Chinese Academy of Sciences, Huizhou 516000, Guangdong Province, China}}

\begin{document}

\title{\boldmath Fully heavy multiquarks in neural-network quantum states}

\author {Wen-min Li}
\iqm
\moe
\sfim
\author {Zhenyu Zhang} \email{Co-first author}
\iqm
\moe
\sfim
\ihep
\ccast
\author {Qian Wang}
\email{qianwang@m.scnu.edu.cn, corresponding author}
\iqm
\sfim
\scnt
 
\date{\today}

\begin{abstract}

Exotic hadrons beyond the conventional quark model provide a direct window into the dynamics of strong interaction. However, extracting the multiquark spectroscopy has to face  the quantum many-body problem, which is still a theoretical challenge. In this case, diquark-antidiquark model is proposed as an approximation. Although this model can describe the spectroscopy roughly, it cannot describe the detailed dynamics. Furthermore, the methods aiming at dealing with many-body problem, e.g. the Gaussian expansion method and diffusion Monte Carlo, are proposed, but  face severe computational bottlenecks. In this work, we introduce the neural-network quantum state (NNQS) approach to investigate the spectra of fully heavy multiquark states within the nonrelativistic potential quark model. By employing deep neural networks to represent the complex many-body spatial wave function, and constructing the color-spin part exactly from group theory to enforce fermionic antisymmetry, our approach effectively overcomes the dimensionality obstacles inherent in traditional methods. The results are compared with various model calculations, demonstrating that NNQS offers superior accuracy and flexibility, particularly in treating high-dimensional correlations. This work establishes NNQS as a promising tool for exploring the spectroscopy of exotic hadrons.
 
\end{abstract}
\maketitle

% keywords can be removed
% {\it1. Introduction.}
\section{INTRODUCTION}

Hadron spectroscopy is one of the most direct ways to shed light on the dynamics of the strong interaction. Due to color confinement in quantum chromodynamics (QCD), any color‑neutral object is allowed--not only the conventional mesons ($q\bar{q}$) and baryons ($qqq$) described by the quark model~\cite{Gell-Mann:1964ewy,Zweig:1964ruk}. Consequently, the search for exotic hadrons beyond the conventional quark model has become one of the most important frontiers in hadron physics. Since the first discovery of the $X(3872)$ by the Belle Collaboration in 2003~\cite{Belle:2003nnu}, a large number of exotic hadron candidates have been observed experimentally, including the $T^+_{cc}(3875)$ 
~\cite{LHCb:2021auc,LHCb:2021vvq}, 
$Z_c(3900)$~\cite{BESIII:2013ris,Belle:2013yex}, $P_c$~\cite{LHCb:2015yax,LHCb:2016ztz,LHCb:2019kea}, 
$P_{cs}$~\cite{LHCb:2022ogu}, and 
$T_{cc\bar{c}\bar{c}}(6900)$~\cite{LHCb:2020bwg}. Several interpretations have been proposed to understand their internal structures~\cite{Guo:2017jvc,Liu:2019zoy,Chen:2016qju,Esposito:2016noz,Brambilla:2019esw,Chen:2022asf,Guo:2019twa}, among which loosely bound hadronic molecules~\cite{Guo:2017jvc,Liu:2019zoy} and compact multiquark states~\cite{Chen:2016qju,Esposito:2016noz,Richard:2016eis,Ali:2017jda,Liu:2019zoy,Chen:2022asf} are the most intensively discussed. The former, having a larger size, corresponds to a bound state formed by the color‑neutral residual strong force, while the latter, with a smaller size, is a bound state formed directly by the fundamental color force. 

Solving the quantum many-body problem for multiquark states presents significant theoretical challenges. For an $N$-quark system, the SU(3) color degree of freedom makes the system more complex than electron or nucleon systems. In previous studies, physical approximations such as the diquark-antidiquark model have been adopted to reduce the complexity of the system~\cite{Wang:2017jtz,Debastiani:2017msn,Galkin:2023wox, Bedolla:2019zwg,Faustov:2020qfm,Anselmino:1992vg,Richard:2016eis,Esposito:2016noz,Ali:2017jda,Liu:2019zoy,Chen:2022asf}. Although this approximation can significantly reduce the system's degrees of freedom, it artificially restricts the wave function to a specific configuration, ignoring the many-body correlation effects between (anti)quarks. Such configurations in wave function may lead to systematic errors and deviate from physical reality.

To overcome the above restricted configurations problem, more rigorous numerical methods are performed to directly solve multiquark systems. Currently, numerical approaches such as diffusion Monte Carlo (DMC)~\cite{Gordillo:2020sgc,Troyer:2004ge,Gordillo:2024blx} and basis expansion methods, including the Gaussian expansion method (GEM)~\cite{Hiyama:2003cu} and the explicitly correlated Gaussian (ECG)~\cite{Varga:1995dm,Mitroy:2013eom} method, have been widely applied to solve multiquark systems. However, these methods face computational bottlenecks when dealing with multiquark systems. For basis expansion methods, the number of basis grows exponentially with the increasing number of quarks, making the precise solution of complex systems highly challenging. And the DMC method inevitably suffers from the fermion sign problem during imaginary time evolution, severely limiting its applicability to strongly correlated multiquark systems. 
Furthermore, wave function cusps~\cite{Pfau:2020ab} arising from the short-range color Coulomb interaction further increase the difficulty of convergence for traditional basis expansion methods. 

Recently, machine learning techniques have provided a new approach to solving the quantum many-body problem. Motivated by their exceptional capacity to approximate high-dimensional functions~\cite{LeCun:2015pmr}, neural-network quantum states (NNQS) offer a flexible wave function representation capable of capturing complex many-body correlations efficiently~\cite{Carleo:2019ptp}. This neural network based variational Monte Carlo (VMC) framework was implemented in quantum spin systems~\cite{Carleo:2016svm} and condensed matter physics~\cite{Li:2022ab,Kim:2023fwy}, as well as ab-initio
chemistry problems~\cite{Pfau:2020ab,Hermann:2020xqs,Choo:2019vcr}. Furthermore, this method has been effectively extended to nuclear physics~\cite{Adams:2020aax,Keeble:2019bkv,Yang:2022esu,Yang:2022rlw,Yang:2024wsg,Yang:2025mhg}. 
In hadron physics, machine learning techniques have demonstrated exceptional predictive power for the mass spectra and decay widths of exotic hadrons~\cite{Malekhosseini:2024eot,Malekhosseini:2025hyx,Rostami:2025sff,Rostami:2026jyz}. However, directly solving the quantum many-body problem in hadron physics remains in the exploratory stage. Recently, the DeepQuark framework~\cite{Wu:2025wvv} was proposed, which marks the first successful implementation of the deep neural networks (DNNs) based VMC method in multiquark systems. These recent developments imply the feasibility of extending this method to multiquark systems.

In this work, we introduce NNQS approach to investigate the mass spectra of fully heavy multiquark states within the nonrelativistic potential quark model, with a particular focus on its advantages over the GEM.
For the color-spin wave function, we construct it based on group theory and symmetry, ensuring that the trial wave function exactly satisfies the fermionic antisymmetry constraint. For the complex many-body spatial wave function, we entirely employ DNNs for high-dimensional fitting. This approach effectively overcomes the computational bottlenecks faced by traditional methods in solving high-dimensional spatial many-body correlations, thereby providing more accurate theoretical predictions for the mass spectra of fully heavy multiquark systems. This paper is organized as follows. In Sec.~\ref{sec:Hamiltonian}, we introduce the nonrelativistic Hamiltonian of the system. The configurations of the multiquark states are constructed in Sec.~\ref{sec:configuration}.
In Sec.~\ref{sec:Neural-network}, we detail the architecture of the neural-network wave function and the optimization procedure. Finally, the numerical results and discussions are presented in Sec.~\ref{sec:Results}.

\section{Hamiltonian}\label{sec:Hamiltonian}
In this work, we adopt the nonrelativistic quark potential model (NRQPM) to describe fully heavy multiquark states. The Hamiltonian is given by:
\begin{equation}
    H = \sum_{i}^{n}(m_i+ T_i ) + \sum_{i<j}^{n}V_{ij}(r_{ij}),
\end{equation}
where $m_i$ and $T_i$ denote the mass and kinetic energy of the $i$th (anti)quark, respectively. The term $V_{ij}(r_{ij})$ represents the effective potential between the $i$th and $j$th (anti)quarks, which depends on their relative distance $r_{ij} = |\bm{r}_i - \bm{r}_j|$. The effective potential $V_{ij}(r_{ij})$ consists of a one-gluon exchange (OGE) potential and a linear confinement potential~\cite{Liu:2019zuc,Liang:2024met}, written as $V_{ij}(r_{ij}) = V_{ij}^{\text{OGE}}(r_{ij}) + V_{ij}^{\text{Conf}}(r_{ij})$. The forms of $V_{ij}^{\text{OGE}}$ and $V_{ij}^{\text{Conf}}$ are given by,
\begin{align}
    V_{ij}^{\text{OGE}} =& \frac{\alpha_{ij}}{4}(\bm{\lambda}_i \cdot \bm{\lambda}_j)\left [ \frac{1}{r_{ij}}-\frac{\pi}{2} \frac{\sigma_{ij}^{3}e^{-\sigma_{ij}^{2}r_{ij}^2}}{\pi^{3/2}}\frac{4}{3m_i m_j}(\bm{\sigma}_i\cdot\bm{\sigma}_j) \right ], \nonumber\\
    V_{ij}^{\text{Conf}}=&-\frac{3}{16}(\bm{\lambda}_i \cdot \bm{\lambda}_j)  br_{ij},
    \label{eq:potential}
\end{align}
where $\bm{\lambda}_i$ is the Gell-Mann matrix acting on the $i$th quark (replaced by $-\bm{\lambda}^{*}$ for antiquarks), and $\bm{\sigma}_i$ represents the Pauli matrix.
The parameters $\alpha_{ij}$ and $\sigma_{ij}$ are the strong coupling strength and the regulator of the spin-spin interaction, respectively. $b$ is the strength of the linear confinement potential. The parameters of model are given in Table~\ref{tab:parameters}, which are well determined by fitting the mass spectra of charmonium, bottomonium, and $B_c$ mesons~\cite{Liu:2019zuc,Liang:2024met}. The parameters and the potential are identical to those used in the GEM of Refs.~\cite{Liu:2019zuc,Liang:2024met,Wang:2026gch}.
\begin{table}[htbp]
\setlength{\tabcolsep}{7mm}
\renewcommand{\arraystretch}{1.5}
\centering
\caption{Parameters of model used in this work~\cite{Liu:2019zuc,Liang:2024met}.}
\label{tab:parameters}
\begin{tabular}{lc}   
\hline
\hline
%\multicolumn{2}{c}{Parameters of model} \\
%\hline
$m_c/m_b$ (GeV)   & 1.483/4.852 \\
$\alpha_{cc}/\alpha_{bc}/\alpha_{bb}$   &  0.5461/0.5021/0.4311\\
$\sigma_{cc}/\sigma_{bc}/\sigma_{bb}$ (GeV)    &  1.1384/1.3000/2.3200\\
$b$ (GeV$^2$)    &  0.1425\\
\hline
\hline
\end{tabular}
\end{table}

\section{The configuration of multiquarks}\label{sec:configuration}
The total wave function of multiquark can be expressed as a direct product of the flavor, spatial, spin, and color wave functions:
\begin{equation}
    \Psi_{\text{total}} = \Psi_{\text{spatial}}\otimes \Psi_{\text{spin}}\otimes\Psi_{\text{color}}\otimes\Psi_{\text{flavor}}.
\end{equation}
In this work, the spin and color wave functions are constructed via $\text{SU(3)} \otimes \text{SU(2)}$ symmetry, and the spatial wave function is parametrized by DNNs. Additionally, by considering only multiquark with identical flavors, the flavor wave function is symmetric.

For conventional mesons and baryons, the construction of their color-spin configurations is well established. For multiquark, the configurations become more complex. The absence of one-light-meson exchanges makes fully heavy systems highly favorable to form genuine compact multiquark configurations rather than conventional hadronic molecules~\cite{Liu:2019zuc,Liang:2024met}. To describe such compact states, we adopt the $\{Q_1Q_2\}\{\bar{Q}_3\bar{Q}_4\}$ configuration for fully heavy tetraquarks~\footnote{Note that this does not mean a diquark-antidiquark approximation, but a way to combine to a color neutral object.}, which includes the following systems: $cc\bar{c}\bar{c}$, $bb\bar{b}\bar{b}$, and $bb\bar{c}\bar{c}$. In the color space, there exist two color-singlet bases for the tetraquark system, whose explicit forms are:
\begin{align}
    \mathcal{C}_1= & \left|(Q_1Q_2)^6(\bar{Q}_3\bar{Q}_4)^{\bar{6}}\right\rangle,\\
    \mathcal{C}_2= & \left|(Q_1Q_2)^{\bar{3}}(\bar{Q}_3\bar{Q}_4)^3\right\rangle.
    \label{eq:singlet}
\end{align}
Here, the superscripts denote the irreducible representations of the $\text{SU(3)}$ color group for the diquark and antidiquark, respectively. Based on the $\mathrm{SU}(3)$ Clebsch-Gordan(CG) coefficients, the explicit forms of the color wave functions can be written as follows~\cite{Liu:2019zuc,Liu:2004kd,Liu:2016ogz,deSwart:1963pdg,Kaeding:1995vq}:
\begin{align}
    \mathcal{C}_1= & \frac{1}{2\sqrt{6}} [ (rb + br)(\bar{b}\bar{r} + \bar{r}\bar{b}) + (gr + rg)(\bar{g}\bar{r} + \bar{r}\bar{g})\nonumber \\
           & + (gb + bg)(\bar{b}\bar{g} + \bar{g}\bar{b}) + 2(rr)(\bar{r}\bar{r}) + 2(gg)(\bar{g}\bar{g}) \nonumber \\
           & + 2(bb)(\bar{b}\bar{b}) ],\\
    \mathcal{C}_2 = &\frac{1}{2\sqrt{3}} [ (br - rb)(\bar{b}\bar{r} - \bar{r}\bar{b}) - (rg - gr)(\bar{g}\bar{r} - \bar{r}\bar{g}) \nonumber\\
           & + (bg - gb)(\bar{b}\bar{g} - \bar{g}\bar{b}) ].
\end{align}
With the color wave functions, the matrix elements $\langle \bm{\lambda}_i \cdot \bm{\lambda}_j \rangle$ can be evaluated~\cite{Vijande:2009ac}, and the results are summarized in Table~\ref{tab:tetraquarks Color interaction}. 
%It is worth mentioning that, according to the GEM in 2019 study~\cite{Liu:2019zuc}, the contribution of the off-diagonal elements can be ignored. For simplicity, we only consider the diagonal matrix elements results in this work to compare the two methods.

\begin{table}[htbp]
\setlength{\tabcolsep}{0.1mm} 
\renewcommand{\arraystretch}{2}
\centering
\caption{Color matrix elements of the tetraquarks.
%The Off-diagonal cross terms are omitted, as previous calculations demonstrate their contributions to the mass are negligible~\cite{Liu:2019zuc}.
}
\label{tab:tetraquarks Color interaction}
\begin{tabular}{p{1.2cm}<{\centering}p{1.2cm}<{\centering}p{1.2cm}<{\centering}p{1.2cm}<{\centering}p{1.2cm}<{\centering}p{1.2cm}<{\centering}p{1.2cm}<{\centering}}     
\hline
\hline
$\hat{O}$& $\bm{\lambda}_1 \cdot\bm{\lambda}_2$ & $\bm{\lambda}_1 \cdot\bm{\lambda}_3$ & $\bm{\lambda}_1 \cdot\bm{\lambda}_4$ & $\bm{\lambda}_2 \cdot\bm{\lambda}_3$ & $\bm{\lambda}_2 \cdot\bm{\lambda}_4$ & $\bm{\lambda}_3 \cdot\bm{\lambda}_4$ \\
\hline
$\langle\mathcal{C}_1| \hat{O}| \mathcal{C}_1  \rangle$ & 4/3 & -10/3 & -10/3 & -10/3 & -10/3 & 4/3 \\
$\langle\mathcal{C}_2| \hat{O}| \mathcal{C}_2  \rangle$ & -8/3 & -4/3 & -4/3 & -4/3 & -4/3 & -8/3 \\
$\langle\mathcal{C}_1| \hat{O}| \mathcal{C}_2  \rangle$ & 0 & $-2\sqrt{2}$ & $2\sqrt{2}$ & $2\sqrt{2}$ & $-2\sqrt{2}$ & 0 \\
\hline
\hline
\end{tabular}
\end{table}

In the spin space, the total spin of the system can take values of $J=0$, $1$, and $2$, are represented as follows:
\begin{align}
    \mathcal{S}_1 &= |(Q_1 Q_2)_0 (\bar{Q}_3 \bar{Q}_4)_0 \rangle_0, \\[2ex]
    \mathcal{S}_2 &= |(Q_1 Q_2)_1 (\bar{Q}_3 \bar{Q}_4)_1 \rangle_0, \\[2ex]
    \mathcal{S}_3 &= |(Q_1 Q_2)_0 (\bar{Q}_3 \bar{Q}_4)_1 \rangle_1, \\[2ex]
    \mathcal{S}_4 &= |(Q_1 Q_2)_1 (\bar{Q}_3 \bar{Q}_4)_0 \rangle_1, \\[2ex]
    \mathcal{S}_5 &= |(Q_1 Q_2)_1 (\bar{Q}_3 \bar{Q}_4)_1 \rangle_1, \\[2ex]
    \mathcal{S}_6 &= |(Q_1 Q_2)_1 (\bar{Q}_3 \bar{Q}_4)_1 \rangle_2.
\end{align}
Here, the subscript denotes spin quantum number. Based on the $\mathrm{SU}(2)$ CG coefficients, the explicit forms of the spin wave functions can be written as follows~\cite{Liu:2019zuc}:
\begin{align}
    \mathcal{S}_1 &= \frac{1}{2} (\uparrow\downarrow\uparrow\downarrow - \uparrow\downarrow\downarrow\uparrow - \downarrow\uparrow\uparrow\downarrow + \downarrow\uparrow\downarrow\uparrow), \\[1ex]
    \mathcal{S}_2 &= \sqrt{\frac{1}{12}} (2\uparrow\uparrow\downarrow\downarrow - \uparrow\downarrow\uparrow\downarrow - \uparrow\downarrow\downarrow\uparrow \notag \\
                   &\quad - \downarrow\uparrow\uparrow\downarrow - \downarrow\uparrow\downarrow\uparrow + 2\downarrow\downarrow\uparrow\uparrow), \\[1ex]
    \mathcal{S}_3 &= \sqrt{\frac{1}{2}} (\uparrow\downarrow\uparrow\uparrow - \downarrow\uparrow\uparrow\uparrow), \\[1ex]
    \mathcal{S}_4 &= \sqrt{\frac{1}{2}} (\uparrow\uparrow\uparrow\downarrow - \uparrow\uparrow\downarrow\uparrow), \\[1ex]
    \mathcal{S}_5 &= \frac{1}{2} (\uparrow\uparrow\uparrow\downarrow + \uparrow\uparrow\downarrow\uparrow - \uparrow\downarrow\uparrow\uparrow - \downarrow\uparrow\uparrow\uparrow), \\[1ex]
    \mathcal{S}_6 &= \uparrow\uparrow\uparrow\uparrow.
\end{align}
With the spin wave functions, the matrix elements $\left \langle\bm{\sigma}_i\cdot\bm{\sigma}_j\right \rangle $ can be evaluated~\cite{Vijande:2009ac}, and the results are summarized
in Table~\ref{tab:Spin matrix elements of the tetraquarks}.
\begin{table}[htbp]
\setlength{\tabcolsep}{0mm} 
\renewcommand{\arraystretch}{2}
\centering
\caption{Spin matrix elements of the tetraquarks. }%The Off-diagonal cross terms are omitted, as previous calculations demonstrate their contributions to the mass are negligible ~\cite{Liu:2019zuc}.}
\label{tab:Spin matrix elements of the tetraquarks}
\begin{tabular}{p{1.2cm}<{\centering}p{1.2cm}<{\centering}p{1.2cm}<{\centering}p{1.2cm}<{\centering}p{1.2cm}<{\centering}p{1.2cm}<{\centering}p{1.2cm}<{\centering}}   
\hline
\hline
$\hat{O}$& $\bm{\sigma}_1 \cdot\bm{\sigma}_2$ & $\bm{\sigma}_1 \cdot\bm{\sigma}_3$ & $\bm{\sigma}_1 \cdot\bm{\sigma}_4$ & $\bm{\sigma}_2 \cdot\bm{\sigma}_3$ & $\bm{\sigma}_2 \cdot\bm{\sigma}_4$ & $\bm{\sigma}_3 \cdot\bm{\sigma}_4$ \\
\hline
$\langle\mathcal{S}_1|\hat{O}|\mathcal{S}_1\rangle$ & -3 & 0 & 0 & 0 & 0 & -3 \\
$\langle\mathcal{S}_2|\hat{O}|\mathcal{S}_2\rangle$ & 1 & -2 & -2 & -2 & -2 & 1 \\
$\langle\mathcal{S}_1|\hat{O}|\mathcal{S}_2\rangle$ & 0 & $-\sqrt{3}$ & $\sqrt{3}$ & $\sqrt{3}$ & $-\sqrt{3}$ & 0 \\
$\langle\mathcal{S}_3|\hat{O}|\mathcal{S}_3\rangle$ & -3 & 0 & 0 & 0 & 0 & 1 \\
$\langle\mathcal{S}_4|\hat{O}|\mathcal{S}_4\rangle$ & 1 & 0 & 0 & 0 & 0 & -3 \\
$\langle\mathcal{S}_5|\hat{O}|\mathcal{S}_5\rangle$ & 1 & -1 & -1 & -1 & -1 & 1 \\
$\langle\mathcal{S}_6|\hat{O}|\mathcal{S}_6\rangle$ & 1 & 1 & 1 & 1 & 1 & 1 \\
\hline
\hline
\end{tabular}
\end{table}

For simplicity, we consider the $\{Q_1Q_2Q_3Q_4\}\bar{Q}_5$ configuration for fully heavy pentaquarks, where the four quarks are identical flavors, which includes the following systems: $cccc\bar{c}$, $cccc\bar{b}$, $bbbb\bar{b}$, and $bbbb\bar{c}$. 
In the color space, three color-singlet states can be constructed via the SU(3) group theory. Its Young tableau representation is given by~\cite{Park:2017jbn}: 
\begin{equation}
C_1 = {\begin{array}{|c|c|}
\hline 1 & 4 \\ \hline
2 & \multicolumn{1}{c}{} \\ \cline{1-1}
3 & \multicolumn{1}{c}{} \\ \cline{1-1}
\end{array}}_{\;3} \otimes (5)_{\bar{3}},~
C_2 = {\begin{array}{|c|c|}
\hline 1 & 2 \\ \hline
3 & \multicolumn{1}{c}{} \\ \cline{1-1}
4 & \multicolumn{1}{c}{} \\ \cline{1-1}
\end{array}}_{\;3} \otimes (5)_{\bar{3}},~
C_3 = {\begin{array}{|c|c|}
\hline 1 & 3 \\ \hline
2 & \multicolumn{1}{c}{} \\ \cline{1-1}
4 & \multicolumn{1}{c}{} \\ \cline{1-1}
\end{array}}_{\;3} \otimes (5)_{\bar{3}}.
\end{equation}
Using the $\mathrm{SU}(3)$ CG coefficients~\cite{Kaeding:1995vq}, the explicit forms of the color wave functions for the pentaquark can be written as follows~\cite{Zhang:2023hmg}:
\begin{align}
C_1 &= \frac{1}{3\sqrt{2}}  (grbb - rgbb + rbgb - brgb + bgrb - gbrb)\bar{b}\nonumber \\
             &\quad + (grbr - rgbr + rbgr - brgr + bgrr - gbrr)\bar{r} \nonumber\\
             &\quad + (grbg - rgbg + rbgg - brgg + bgrg - gbrg)\bar{g},\\
C_2 &= \frac{1}{4\sqrt{3}} (2bbgr - 2bbrg + gbrb - gbbr + bgrb - bgbr \nonumber\\
             &\quad - rbgb + rbbg - brgb + brbg)\bar{b} + (2rrbg - 2rrgb \nonumber\\
             &\quad + rgrb - rgbr + grrb - grbr + rbgr - rbrg + brgr \nonumber\\
             &\quad - brrg)\bar{r} + (2ggrb - 2ggbr - rggb + rgbg - grgb \nonumber\\
             &\quad + grbg + gbgr - gbrg + bggr - bgrg)\bar{g}],\\
C_3 &= \frac{1}{12} (3bgbr - 3gbbr - 3brbg + 3rbbg - rbgb - 2rgbb \nonumber\\
             &\quad + 2grbb + brgb + gbrb - bgrb)\bar{b} + (3grrb - 3rgrb\nonumber \\
             &\quad - 3brrg + 3rbrg - rbgr - 2gbrr + 2bgrr - grbr\nonumber \\
             &\quad + rgbr + brgr)\bar{r} + (3grgb - 3rggb + 3bggr - 3gbgr \nonumber\\
             &\quad - grbg + rgbg + 2rbgg - 2brgg + gbrg - bgrg)\bar{g}.
\end{align}

In the spin space, based on the $\text{SU(2)}$ coupling rules, the total spin of the system can take values of $J=5/2, 3/2$, and $1/2$, and their corresponding Young tableaux are represented as follows:
\begin{align}
    \ytableausetup{boxsize=1.2em}
    J = \frac{5}{2}: \quad & S_1 = \begin{ytableau} 1 & 2 & 3 & 4 & 5 \end{ytableau} . \\[3ex]
    J = \frac{3}{2}: \quad & S_2 = \begin{ytableau} 1 & 2 & 3 & 4 \\ 5 \end{ytableau} , \,
                             S_3 = \begin{ytableau} 1 & 2 & 3 & 5 \\ 4 \end{ytableau} , \notag \\
                           & S_4 = \begin{ytableau} 1 & 3 & 4 & 5 \\ 2 \end{ytableau} , \,
                             S_5 = \begin{ytableau} 1 & 2 & 4 & 5 \\ 3 \end{ytableau} . \\[3ex]
    J = \frac{1}{2}: \quad & S_6 = \begin{ytableau} 1 & 2 & 3 \\ 4 & 5 \end{ytableau} , \,
                             S_7 = \begin{ytableau} 1 & 3 & 4 \\ 2 & 5 \end{ytableau} , \,
                             S_8 = \begin{ytableau} 1 & 2 & 4 \\ 3 & 5 \end{ytableau} , \notag \\
                           & S_9 = \begin{ytableau} 1 & 2 & 5 \\ 3 & 4 \end{ytableau} , \,
                             S_{10} = \begin{ytableau} 1 & 3 & 5 \\ 2 & 4 \end{ytableau} .
\end{align}
Incorporating the $\mathrm{SU}(2)$ CG coefficients, the explicit forms of the pentaquark spin wave functions can be derived:
\begin{align}
    S_1 =&\uparrow \uparrow \uparrow \uparrow \uparrow,  \\
    S_2 =&\frac{1}{2\sqrt{5}}(4\uparrow \uparrow \uparrow \uparrow \downarrow -\uparrow \uparrow \uparrow \downarrow \uparrow -\uparrow \uparrow \downarrow \uparrow \uparrow -\uparrow \downarrow \uparrow \uparrow \uparrow -\downarrow \uparrow \uparrow \uparrow \uparrow),\\
    S_3 =& \frac{1}{2\sqrt{3}} (3\uparrow\uparrow\uparrow\downarrow\uparrow - \uparrow\uparrow\downarrow\uparrow\uparrow - \uparrow\downarrow\uparrow\uparrow\uparrow - \downarrow\uparrow\uparrow\uparrow\uparrow), \\
    S_4 =& \frac{1}{\sqrt{2}} (\uparrow\downarrow\uparrow\uparrow\uparrow - \downarrow\uparrow\uparrow\uparrow\uparrow), \\
    S_5 =& \frac{1}{\sqrt{6}} (2\uparrow\uparrow\downarrow\uparrow\uparrow - \uparrow\downarrow\uparrow\uparrow\uparrow - \downarrow\uparrow\uparrow\uparrow\uparrow), 
\end{align}
\begin{align}
    S_6 =& \frac{1}{3\sqrt{2}} \Bigl( 3\uparrow\uparrow\uparrow\downarrow\downarrow - \uparrow\uparrow\downarrow\uparrow\downarrow - \uparrow\downarrow\uparrow\uparrow\downarrow - \downarrow\uparrow\uparrow\uparrow\downarrow - \uparrow\uparrow\downarrow\downarrow\uparrow \nonumber \\
    & - \uparrow\downarrow\uparrow\downarrow\uparrow - \downarrow\uparrow\uparrow\downarrow\uparrow + \uparrow\downarrow\downarrow\uparrow\uparrow + \downarrow\uparrow\downarrow\uparrow\uparrow + \downarrow\downarrow\uparrow\uparrow\uparrow \Bigr), \\
    S_7 =& \frac{1}{2\sqrt{3}} (2\uparrow\downarrow\uparrow\uparrow\downarrow - 2\downarrow\uparrow\uparrow\uparrow\downarrow - \uparrow\downarrow\downarrow\uparrow\uparrow+ \downarrow\uparrow\downarrow\uparrow\uparrow - \uparrow\downarrow\uparrow\downarrow\uparrow\nonumber \\
    &  + \downarrow\uparrow\uparrow\downarrow\uparrow), \\
    S_8 =& \frac{1}{6} \Bigl( 4\uparrow\uparrow\downarrow\uparrow\downarrow - 2\uparrow\downarrow\uparrow\uparrow\downarrow - 2\downarrow\uparrow\uparrow\uparrow\downarrow - 2\uparrow\uparrow\downarrow\downarrow\uparrow+ 2\downarrow\downarrow\uparrow\uparrow\uparrow  \nonumber \\
    & + \uparrow\downarrow\uparrow\downarrow\uparrow + \downarrow\uparrow\uparrow\downarrow\uparrow - \uparrow\downarrow\downarrow\uparrow\uparrow - \downarrow\uparrow\downarrow\uparrow\uparrow \Bigr), \\
    S_9 =& \frac{1}{2\sqrt{3}} (2\uparrow\uparrow\downarrow\downarrow\uparrow + 2\downarrow\downarrow\uparrow\uparrow\uparrow - \uparrow\downarrow\uparrow\downarrow\uparrow- \downarrow\uparrow\uparrow\downarrow\uparrow - \uparrow\downarrow\downarrow\uparrow\uparrow\nonumber\\
    &  - \downarrow\uparrow\downarrow\uparrow\uparrow),\\
    S_{10} =&\frac{1}{2}(\uparrow \downarrow \uparrow \downarrow \uparrow -\uparrow \downarrow \downarrow \uparrow \uparrow -\downarrow \uparrow \uparrow \downarrow \uparrow +\downarrow \uparrow \downarrow \uparrow \uparrow ). 
\end{align}

By combining the color and spin wave functions with the CG coefficients of the $S_4$ permutation group~\cite{Stancu:1999qr}, the color-spin wave functions $\Psi^{\text{CS}}$ can be constructed as follows. For the $S$-wave pentaquark system with the $\{Q_1Q_2Q_3Q_4\}\bar{Q}_5$ configuration, there are two  wave functions corresponding to $J^P = 3/2^-$ and $1/2^-$:
\begin{align}
    \Psi^{\text{CS}}_{3/2^-} &= \frac{1}{\sqrt{3}} (C_1 S_3 + C_2 S_4 - C_3 S_5), \\[1ex]
    \Psi^{\text{CS}}_{1/2^-} &= \frac{1}{\sqrt{3}} (C_1 S_6 + C_2 S_7 - C_3 S_8).
    \label{eq:CSwavefunction}
\end{align}
One notices that there is no $J^P=5/2^-$ pentaquark in this configuration, as it violates Fermi-Dirac statistics.
It is worth mentioning that since both the spatial and flavor wave functions of the $\{Q_1Q_2Q_3Q_4\}$ identical quarks are symmetric, the Pauli principle dictates that their color-spin wave function must be totally antisymmetric. States with a total spin of $J^P=5/2^-$ are forbidden for such pentaquark configurations.
\begin{table}[htbp]
\setlength{\tabcolsep}{1mm}
\renewcommand{\arraystretch}{2.5}
\centering
\caption{Color and spin interaction for the pentaquarks.}
\label{tab:Color and color-spin interaction for the pentaquarks}
\begin{tabular}{lccc}   
\hline
\hline
$(i,j)$& $\left\langle\bm\lambda_i\cdot \bm\lambda_j  \right \rangle$   & $\left\langle(\bm\lambda_i\cdot \bm\lambda_j)(\bm\sigma_i\cdot \bm\sigma_j)  \right \rangle_{\frac{3}{2}^{-}}$ & $\left\langle(\bm\lambda_i\cdot \bm\lambda_j)(\bm\sigma_i\cdot \bm\sigma_j)\right \rangle_{\frac{1}{2}^{-}}$\\
\hline
$(1,2)$ &-4/3 &-28/9&-28/9\\
$(1,3)$ &-4/3 &-28/9&-28/9\\
$(1,4)$ &-4/3 &-28/9&-28/9\\
$(1,5)$ &-4/3 &4/3  &-8/3\\
$(2,3)$ &-4/3 &-28/9&-28/9\\
$(2,4)$ &-4/3 &-28/9&-28/9\\
$(2,5)$ &-4/3 &4/3  &-8/3\\
$(3,4)$ &-4/3 &-28/9&-28/9\\
$(3,5)$ &-4/3 &4/3  &-8/3\\
$(4,5)$ &-4/3 &4/3  &-8/3\\
\hline
\hline
\end{tabular}
\end{table}

Based on the color-spin wave functions, one can evaluate the interactions within pentaquarks. Because the potential of $V^{\text{Conf}}$ term depends on color interactions, it yields only diagonal matrix elements due to the orthogonality of the wave functions. The term $V^{\text{OGE}}$ contains color-spin interactions, both diagonal and off-diagonal elements are evaluated. The calculated results are summarized in Table~\ref{tab:Color and color-spin interaction for the pentaquarks}, where the subscript of the color-spin interactions, $\left\langle\cdots\right \rangle_{\frac{3}{2}^{-}}$ and $\left\langle\cdots\right \rangle_{\frac{1}{2}^{-}}$, indicate that the operator acts on different color-spin wave functions.

 \section{Neural-network wave function}\label{sec:Neural-network}
 For the spatial wave function of multiquarks, it is determined by solving the many-body Schr\"{o}dinger equation. Exactly solving the Schr\"{o}dinger equation for multiquarks is a high-dimensional quantum many-body problem. To investigate the ground state properties of the system, we employ the variational method,  the parameters of the trial wave function $\Psi_{\theta}$ are optimized by minimizing the value of energy expectation.
 \begin{equation}
     E_{\theta}=\frac{\left \langle \Psi_{\theta } \right |H\left | \Psi_{\theta }  \right \rangle  }{\left \langle \Psi_{\theta }  | \Psi_{\theta }  \right \rangle }\ge E_0.
 \end{equation}
We employ DNNs to parametrize $\Psi_{\text{spatial}}$. We introduce the three-dimensional coordinates of all quarks $\bm{r}_1, \bm{r}_2, \dots, \bm{r}_n$, which constitute a $3n$-dimensional vector. To remove the kinetic energy contribution from the center of mass motion~\cite{Massella:2018xdj}, we define the internal coordinates $\bar{\bm{r}}_1, \bar{\bm{r}}_2, \dots, \bar{\bm{r}}_n$ as the neural network input, where $\bar{\bm{r}_i}=\bm{r}_i-\bm{R}_{\text{CM}}$, $\bm{R}_{\text{CM}}$ is the center of mass coordinate of the system. The scalar output of the network is $\ln(\Psi_{\text{spatial}})$. This logarithmic representation is adopted to prevent numerical instabilities during the Monte Carlo sampling. To ensure the symmetry of the spatial wave function, we symmetrize the neural network wave function. Furthermore, we added a Gaussian function $-\alpha \sum_{i=1}^n |\bar{\bm{r}}_i|^2$ to confine the particle within a finite volume~\cite{Adams:2020aax}, where the parameter $\alpha = 0.05 \text{ fm}^{-2}$.

We employ a fully connected neural network for this work. Each hidden layer applies a linear function to its inputs, followed by a nonlinear activation function:
\begin{equation}
    \bar{\bm{r}}^{(l+1)} = f(\bm{\omega}^{(l)}\bar{\bm{r}}^{(l)}+b^{(l)}),
\end{equation}
where $\bm{\omega}^{(l)}$ and $b^{(l)}$ denote the weights and biases of the $l$th layer, respectively, and $f$ is the activation function. We employ the Sigmoid linear unit (SiLU) as the activation function in this work. Its specific form is expressed as:
\begin{equation}
    \text{SiLU}(x) = \frac{x}{1 + e^{-x}}.
\end{equation}
For meson and baryon systems, the network consists of 4 hidden layers with 24 nodes per layer. For tetraquark and pentaquark systems, the network has 6 hidden layers with 32 nodes each. This design ensures sufficient expressive power while preserving computational feasibility. See Appendix~\ref{sup:neural-network} for more details.

To obtain the optimal parameters $\theta$ that minimize the energy expectation value $E_{\theta}$, we employ the stochastic reconfiguration (SR) algorithm to optimize the variational parameters~\cite{Sorella:1998him,Sorella:2005wav}. The SR algorithm is equivalent to performing imaginary-time evolution in the variational manifold and is closely related to the natural gradient descent method in unsupervised learning~\cite{Adams:2020aax,Amari:1998nat}. Compared to standard gradient descent, the SR approach enhances optimization stability. 
Specifically, at each iteration, we use Metropolis-Hastings Monte Carlo sampling~\cite{Metropolis:1953am,Hastings:1970aa} generates a large coordinate samples, which distributed as $|\Psi_{\theta}|^2$. During the sampling process, a cutoff is introduced to resolve the issue of potential singularities in Eq.~\eqref{eq:potential}.  Therefore, no cusp condition is implemented in the present algorithm. Furthermore, the neural-network wave function includes contributions from all orbital angular momentum, the variational method leads the wave function to converge to the $S$-wave ground state in the optimization process~\cite{Wu:2025wvv}. From these samples, the energy expectation $E_{\theta}$ and its gradients $\nabla_{\theta}E_{\theta}$ are evaluated. The $\nabla_{\theta}E_{\theta}$ is calculated using the following formula:
\begin{equation}
    \nabla_{\theta}E_{\theta}=2\left(\frac{\left\langle\partial {\psi_{\theta }}\right|H\left | \psi_{\theta }  \right \rangle  }{\left \langle \psi_{\theta }  | \psi_{\theta }  \right \rangle }-E_{\theta}\frac{\left \langle \partial \psi_{\theta }  | \psi_{\theta }  \right \rangle }{\left \langle \psi_{\theta }  | \psi_{\theta }  \right \rangle }\right ),
\end{equation}
where $|\partial \psi_\theta\rangle \equiv \partial |\psi_\theta\rangle / \partial \theta$, $S$ matrix denotes the quantum Fisher information matrix, which is defined as:
\begin{equation}
    S_{ab} = \frac{\langle \partial_a \psi_\theta | \partial_b \psi_\theta \rangle}{\langle \psi_\theta | \psi_\theta \rangle} - \frac{\langle \partial_a \psi_\theta | \psi_\theta \rangle \langle \psi_\theta | \partial_b \psi_\theta \rangle}{\langle \psi_\theta | \psi_\theta \rangle^2}.
\end{equation}
%The SR algorithm is equivalent to performing imaginary-time evolution in the variational manifold and is closely related to the natural gradient descent method in unsupervised learning~\cite{Adams:2020aax,amari1998natural}. Compared to standard gradient descent, the SR approach enhances optimization stability.

By calculating  $\nabla_{\theta}E_{\theta}$ and the $S$ matrix, the parameters of the neural network can be updated through the parameter update rule, which is given by:
\begin{align}
    \theta^{i+1}=\theta^i-\eta(S+\epsilon I)^{-1}\nabla_{\theta^i}E_{\theta^i},
    \label{Parameter Update Rule}
\end{align}
where $i$ is the iteration step, $\eta$ is the learning rate, $\epsilon$ is a small regularization constant set between $10^{-3}$ and $10^{-4}$ introduced to ensure matrix invertibility and $I$ is the identity matrix. After hundreds of iterations, the energy expectation value converges to a minimum, yielding a neural-network wave function that approximates the ground state.

%To ensure that the neural network efficiently explores the parameter space during the initial training phase and smoothly transitions to fine-tuning in the late stages, we implement a cosine annealing learning rate schedule. 
\begin{figure*}[htbp]
    \centering
    \includegraphics[width=\textwidth]{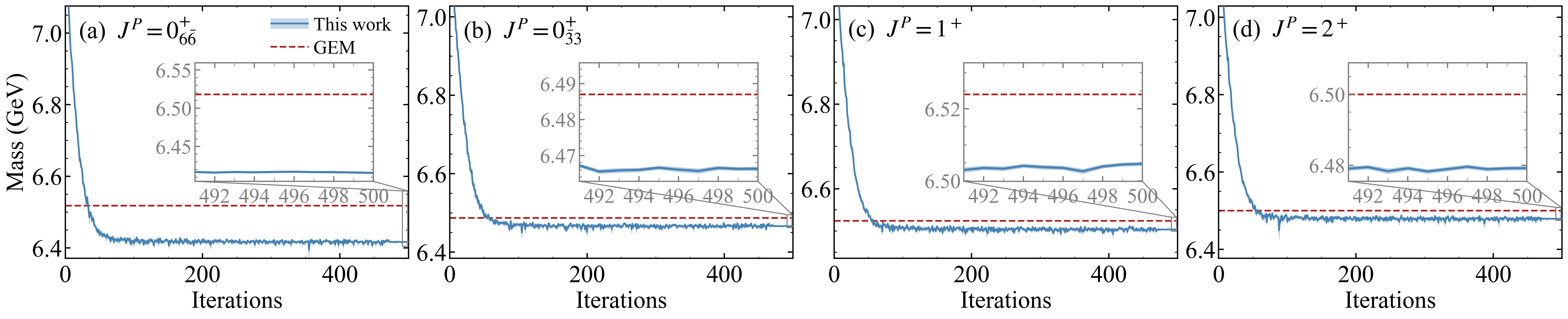}
    \caption{The mass estimate as a function of iteration steps for $cc\bar{c}\bar{c}$ tetraquark configurations (a)$\left|\left \{ cc \right \}_0^6\left \{ \bar{c}\bar{c} \right \}_0^{\bar{6}}\right \rangle _0^0$, (b)$\left |\left \{ cc \right \}_1^{\bar{3}}\left \{ \bar{c}\bar{c} \right \}_1^3   \right \rangle _0^0 $, (c)$\left |\left \{ cc \right \}_1^{\bar{3}}\left \{ \bar{c}\bar{c} \right \}_1^3   \right \rangle _1^0 $ and (d)$\left |\left \{ cc \right \}_1^{\bar{3}}\left \{ \bar{c}\bar{c} \right \}_1^3   \right \rangle _2^0 $ in the optimization progress. The Monte Carlo statistical errors of the masses are shown by the blue shaded area, and red dashed line represents the GEM prediction by Liu et al.~\cite{Liu:2019zuc}. The insets in each subplot magnifies the final optimization stage.}
    \label{fig:NNtetraquark}
\end{figure*}

To accelerate the optimization process, we use a small batch size during the early stage. As optimization enters the final 30 steps, we increase the batch size from 2000 to 80,000 to reduce statistical errors. Throughout iterations, the learning rate decayed from $10^{-1}$ to $10^{-5}$. To describe the random error of the SR algorithm, we employ a multistep averaging scheme in the final stage of optimization. Specifically, we take the average of the masses from the final 10 iterations as our final result. Consequently, the total error $\delta_{\text{tot}}$ is obtained by combining the Monte Carlo statistical error $\delta_{\text{MC}}$ and the neural network error $\delta_{\text{NN}}$, the $\delta_{\text{tot}}$ is given by: 
\begin{equation}
    \delta_{\text{tot}} = \sqrt{\delta_{\text{MC}}^2 + \delta_{\text{NN}}^2},
\end{equation}
where $\delta_{\text{NN}}$ is the standard deviation of the last ten iterations.
All of the above calculations were carried out within the PyTorch~\cite{Paszke:2019pyt} framework.

\section{Results and discussion}\label{sec:Results}
In this section, we present the mass spectra of fully heavy multiquark systems calculated via NNQS. All numerical calculations in this work focus exclusively on the $S$-wave ground states. Before applying the NNQS method to fully heavy multiquark, we first tested it on conventional mesons and baryons. The results are summarized in Table~\ref{tab:The masses of meson and baryon}.
\begin{table}[htbp]
\setlength{\tabcolsep}{4mm}
\renewcommand{\arraystretch}{1.4}
\centering
\caption{The masses (MeV) of meson and baryon. Our results are compared with GEM results by Liu and Ling et al.~\cite{Liu:2019zuc,Liang:2024met} and experimental data~\cite{ParticleDataGroup:2020ssz}.}
\label{tab:The masses of meson and baryon}
\begin{tabular}{lcccc}   
\hline
\hline
State&$J^P$ & This work & GEM & Exp.\\
\hline
$\eta_c$               &$0^-$ & $2983.7(10)$   &2983  &2984\\
$J/\psi$               &$1^-$ & $3097.2(8)$   &3097  &3097\\
$B_c$                  &$0^-$ & $6270.7(9)$   &6271  &6274\\
$B_{c}^{\ast}$         &$1^-$ & $6326.3(10)$   &6328  &$\cdots$\\
$\eta_b$               &$0^-$ & $9389.4(18)$   &9390  &9399\\
$\Upsilon$             &$1^-$ & $9460.0(12)$   &9460  &9460\\
$\Omega_{ccc}$         &$3/2^+$ & $4816.1(5)$   &4823  &$\cdots$\\
$\Omega_{bbb}$         &$3/2^+$ & $14408.0(9)$  &14421 &$\cdots$\\
$\Omega_{ccb}$         &$1/2^+$ & $8025.3(7)$   &8034   &$\cdots$\\
$\Omega_{ccb}^{\ast}$  &$3/2^+$ & $8050.3(6)$   &8057   &$\cdots$\\
$\Omega_{cbb}$         &$1/2^+$ & $11217.6(8)$  &11222  &$\cdots$\\
$\Omega_{cbb}^{\ast}$  &$3/2^+$ & $11246.1(7)$  &11250  &$\cdots$\\
\hline
\hline
\end{tabular}
\end{table}
For meson, the results show that the masses using our method agree with those from the GEM~\cite{Liu:2019zuc,Liang:2024met}, which utilize Gaussian basis functions expanded in standard rearrangement channels as shown in Table~\ref{tab:The masses of meson and baryon}.  This shows that the neural-network wave function is capable of effectively describing the spatial correlations of the two-body system. When the number of (anti)quarks increases to three, i.e. baryons, the superiority of the NNQS method becomes prominent. The masses of baryons in this work are lower about 4 to 13 MeV than those in Ref.~\cite{Liang:2024met}, implying that our variational method produces results very close to the exact values. This suggests that the neural-network wave function has stronger expressive power than trial wave function constructed with predefined Gaussian bases in Ref.~\cite{Liang:2024met}. 
The reason is that the spatial wave function of many-body systems is complex.
The neural network, as a universal function approximator without the need for predefined functional forms, can capture the internal spatial distribution of three-body and more complex systems.

And then, we extend our method to fully heavy tetraquark. The optimization performance of NNQS in tetraquark is shown in Fig.~\ref{fig:NNtetraquark}, and we take different quantum number $cc\bar{c}\bar{c}$ systems as examples. The expected masses for the four configurations decrease rapidly during the initial optimization phase, converging stably below the GEM results presented by Liu et al.~\cite{Liu:2019zuc}. The Monte Carlo statistical uncertainties (blue shaded in Fig.~\ref{fig:NNtetraquark}) are limited and stable, confirming that the NNQS method provides a reliable prediction.
\begin{figure*}[htbp]
    \centering
    \includegraphics[width=\textwidth]{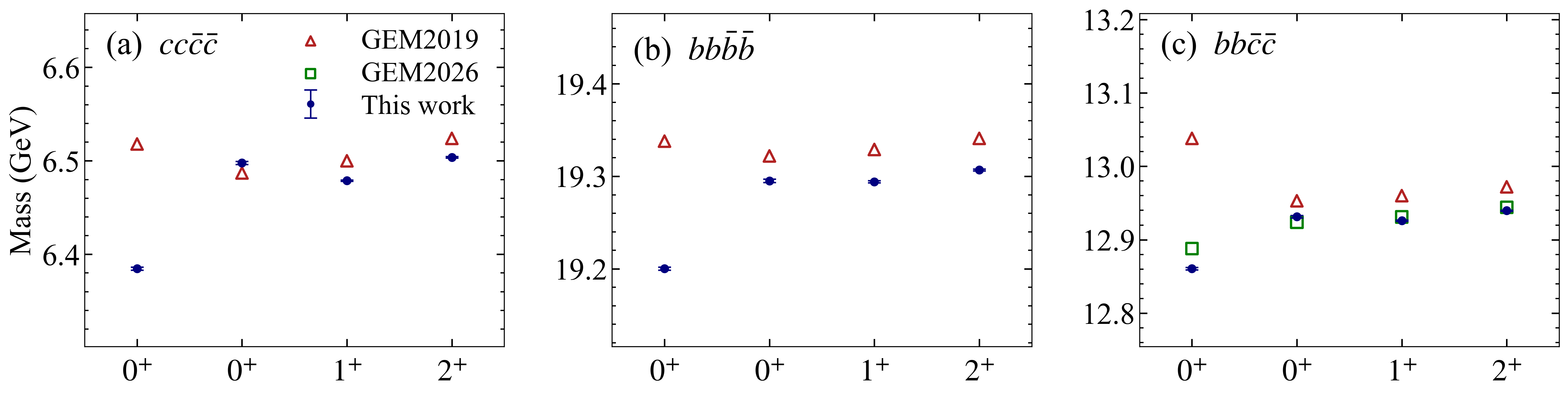}
    \caption{A comparison of the predicted masses of the tetraquark states $cc\bar{c}\bar{c}$ (a), $bb\bar{b}\bar{b}$ (b), and $bb\bar{c}\bar{c}$ (c). The blue points, red triangles, and green squares represent this work predictions, the GEM predictions by Liu et al. in 2019~\cite{Liu:2019zuc} and by Wang et al. in 2026~\cite{Wang:2026gch}, respectively. 
    %The results of this work and GEM are are based on the unmixed $0^{+}_{6\bar{6}}$ and $0^{+}_{\bar{3}3}$ configurations.
    }
    \label{fig:tetraquark comparison with GEM}
\end{figure*}
\begin{table*}[htbp]
\setlength{\tabcolsep}{8mm}
\renewcommand{\arraystretch}{2}
\centering
\caption{Predicted mass (MeV) spectra of tetraquarks for $cc\bar{c}\bar{c}$, $bb\bar{b}\bar{b}$, and $bb\bar{c}\bar{c}$ systems.}
\label{tab:The masses of tetraquarks}
\begin{tabular}{llcccc}   
\hline
\hline
State&$J^{P}$& Configuration & $\langle H \rangle$ & Mass \\
\hline
\multirow{4}{*}{$cc\bar{c}\bar{c}$}
&\multirow{2}{*}{$0^{+}$}  & $\left |\left \{ cc \right \}_0^6\left \{ \bar{c}\bar{c} \right \}_0^{\bar{6}}   \right \rangle _0^0 $   &\multirow{2}{*}{$\begin{pmatrix} 6416.4(8) & -50.7(17) \\ -50.7(17) & 6466.2(7) \end{pmatrix}$} &  6384.8(16) \\
&  & $\left |\left \{ cc \right \}_1^{\bar{3}}\left \{ \bar{c}\bar{c} \right \}_1^3   \right \rangle _0^0 $   & & 6497.8(16) \\
&$1^{+}$             & $\left |\left \{ cc \right \}_1^{\bar{3}}\left \{ \bar{c}\bar{c} \right \}_1^3   \right \rangle _1^0 $   & $\begin{pmatrix} 6479.0(7) \end{pmatrix} $ & 6479.0(7) \\
&$2^{+}$             & $\left |\left \{ cc \right \}_1^{\bar{3}}\left \{ \bar{c}\bar{c} \right \}_1^3   \right \rangle _2^0 $   & $\begin{pmatrix} 6503.8(8) \end{pmatrix}$ & 6503.8(8) \\

\multirow{4}{*}{$bb\bar{b}\bar{b}$}
&\multirow{2}{*}{$0^{+}$}  & $\left |\left \{ bb \right \}_0^6\left \{ \bar{b}\bar{b} \right \}_0^{\bar{6}}   \right \rangle _0^0 $   & \multirow{2}{*}{$\begin{pmatrix} 19209.1(11) & -27.6(23) \\ -27.6(23) & 19286.4(11) \end{pmatrix}$} & 19200.3(17)  \\
&  & $\left |\left \{ bb \right \}_1^{\bar{3}}\left \{ \bar{b}\bar{b} \right \}_1^3   \right \rangle _0^0 $   & & 19295.2(17)\\
&$1^{+}$             & $\left |\left \{ bb \right \}_1^{\bar{3}}\left \{ \bar{b}\bar{b} \right \}_1^3   \right \rangle _1^0 $   & $\begin{pmatrix} 19294.2(12) \end{pmatrix}$ & 19294.2(12) \\
&$2^{+}$             & $\left |\left \{ bb \right \}_1^{\bar{3}}\left \{ \bar{b}\bar{b} \right \}_1^3   \right \rangle _2^0 $   & $\begin{pmatrix} 19306.9(10) \end{pmatrix}$ & 19306.9(10) \\

\multirow{4}{*}{$bb\bar{c}\bar{c}$}
&\multirow{2}{*}{$0^{+}$}  & $\left |\left \{ bb \right \}_0^6\left \{ \bar{c}\bar{c} \right \}_0^{\bar{6}}   \right \rangle _0^0 $   & \multirow{2}{*}{$\begin{pmatrix} 12872.7(8) & -26.5(19) \\ -26.5(19) & 12919.6(7) \end{pmatrix}$} & 12860.8(16) \\
&  & $\left |\left \{ bb \right \}_1^{\bar{3}}\left \{ \bar{c}\bar{c} \right \}_1^3   \right \rangle _0^0 $   & &12931.5(16) \\
&$1^{+}$             & $\left |\left \{ bb \right \}_1^{\bar{3}}\left \{ \bar{c}\bar{c} \right \}_1^3   \right \rangle _1^0 $   & $\begin{pmatrix} 12926.2(9) \end{pmatrix}$ & 12926.2(9) \\
&$2^{+}$             & $\left |\left \{ bb \right \}_1^{\bar{3}}\left \{ \bar{c}\bar{c} \right \}_1^3   \right \rangle _2^0 $   & $\begin{pmatrix} 12939.6(8) \end{pmatrix}$ & 12939.6(8)\\
\hline
\hline
\end{tabular}
\end{table*}
All predicted mass spectra for the $cc\bar{c}\bar{c}$, $bb\bar{b}\bar{b}$, and $bb\bar{c}\bar{c}$ systems have been collected in Table~\ref{tab:The masses of tetraquarks}. We label the two configurations of the $J^P=0^+$ state $\left |\left \{ QQ \right \}_0^6\left \{ \bar{Q}\bar{Q} \right \}_0^{\bar{6}}   \right \rangle _0^0 $ and $\left |\left \{ QQ \right \}_1^{\bar{3}}\left \{ \bar{Q}\bar{Q} \right \}_1^3   \right \rangle _0^0 $ as $0^{+}_{6\bar{6}}$ and $0^{+}_{\bar{3}3}$, respectively. The two physical $0^{+}$ states are a mixture of $0^{+}_{6\bar{6}}$ and $0^{+}_{\bar{3}3}$ configurations.
We first determine the two diagonal elements via neural network training. 
Furthermore, the off-diagonal elements are evaluated by the well trained network, i.e. using the wave functions determined by the training of the diagonal elements. With both diagonal and off-diagonal elements known, one can diagonalize the matrix to extract the mass spectra. We found that the mass of the $0^{+}_{\bar{3}3}$ state is slightly larger than that of $0^{+}_{6\bar{6}}$ state. The mixing further pushes the lower state down and the higher state up, thereby enlarging their mass splitting. In all the tetraquark systems, this splitting between the two physical $J^P=0^+$ states is from 70 to 110 MeV. The other two  $J^P=1^+$ and $J^P=2^+$ fully charm tetraquarks are $6479.0\pm 0.7~\mathrm{MeV}$ and $6503.8\pm 0.8~\mathrm{MeV}$, respectively, significantly lower than the values from Ref.~\cite{Liu:2019zuc}. The mass splitting between these two states is about 25 MeV. 
These masses and the mass splitting indicate that the observed $X(6900)$ cannot be an $S$-wave ground fully charmed tetraquark state. 

\begin{figure*}[htbp]
    \centering
    \includegraphics[width=\textwidth]{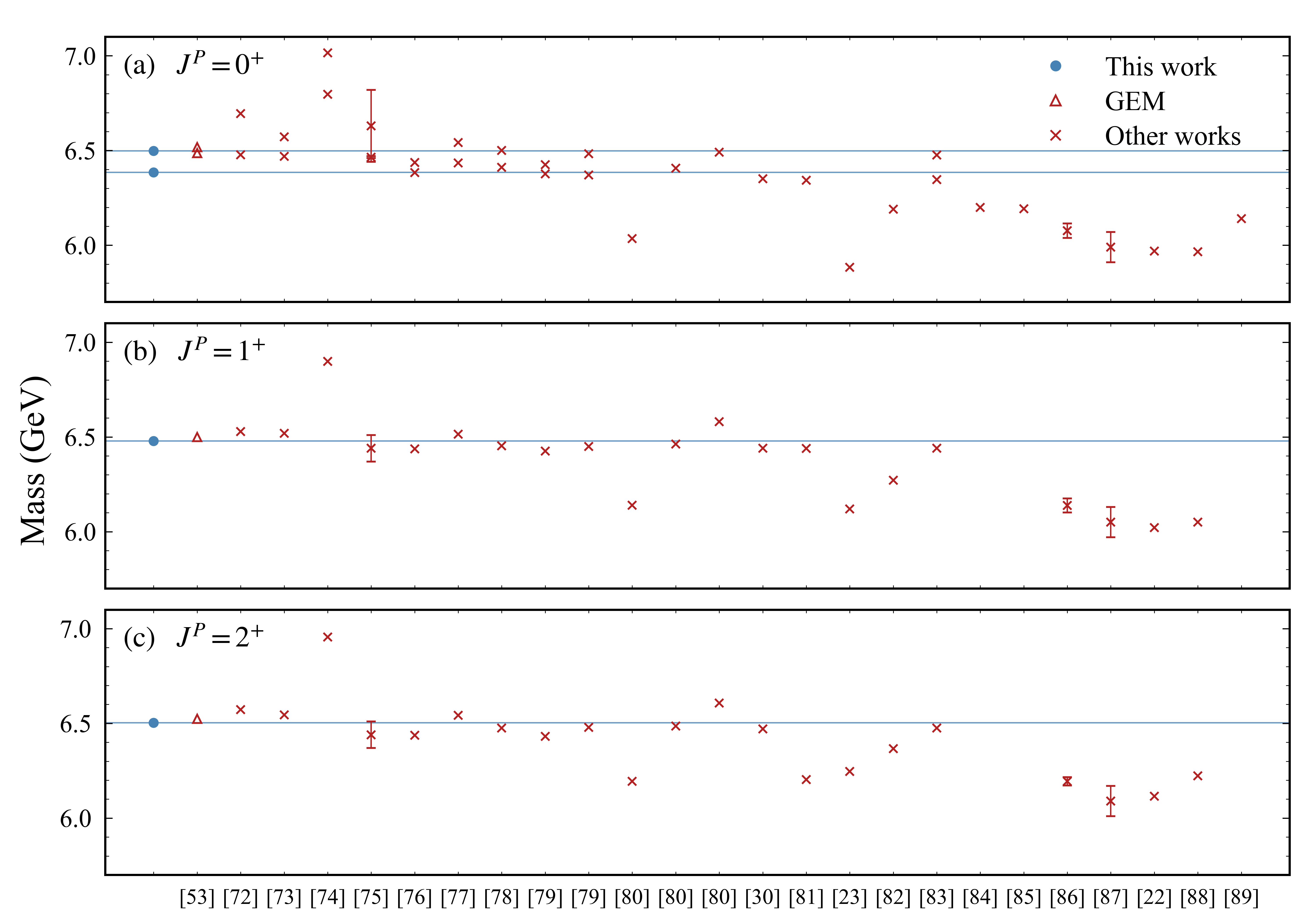}
    \caption{A comparison of the predicted masses of the tetraquark state $cc\bar{c}\bar{c}$ with $J^P=0^+$ (a), $J^P=1^+$ (b), and $J^P=2^+$ (c). The blue points (horizontal lines), red triangles and red crosses represent the predictions from this work, GEM by Liu et al.~\cite{Liu:2019zuc}, and other works, respectively.}
    \label{fig:tetraquark_mass_comparison}
\end{figure*}
\begin{table}[htbp]
\setlength{\tabcolsep}{1mm}
\renewcommand{\arraystretch}{1.5}
\centering
\caption{Our predicted masses(MeV) for the $cc\bar{c}\bar{c}$ system compared with others.}
\label{tab:tetraquark_masses_compared_with_others}
\begin{tabular}{lcccc}   
\hline
\hline
$J^P$ & $0^{+}$  & $0^{+}$  & $1^{+}$ & $2^{+}$ \\
\hline
This work                           & $6384.8(16)$ & $6497.8(16)$ & $6479.0(7)$ & $6503.8(8)$ \\
Ref.~\cite{Liu:2019zuc}              & 6518 & 6487 & 6500 & 6524 \\
Ref.~\cite{Lloyd:2003yc}             & 6695 & 6477 & 6528 & 6573 \\
Ref.~\cite{Yan:2023lvm}              & 6572 & 6469 & 6519 & 6545 \\
Ref.~\cite{Wu:2016vtq}               & 7016 & 6797 & 6899 & 6956 \\
Ref.~\cite{Chen:2016jxd}             & 6440-6820 & 6460-6470 & 6370-6510 & 6370-6510 \\
Ref.~\cite{Ader:1981db}              & 6383 & 6437 & 6437 & 6437 \\
Ref.~\cite{Lu:2020cns}               & 6435 & 6542 & 6515 & 6543 \\
Ref.~\cite{Zhang:2022qtp}            & 6500 & 6411 & 6453 & 6475 \\
Ref.~\cite{Wang:2019rdo}             & 6377 & 6425 & 6425 & 6432 \\
Ref.~\cite{Wang:2019rdo}             & 6371 & 6483 & 6450 & 6479 \\
Ref.~\cite{Deng:2020iqw}             & $\cdots$ & 6035 & 6139 & 6194 \\
Ref.~\cite{Deng:2020iqw}             & 6407 & $\cdots$ & 6463 & 6486 \\
Ref.~\cite{Deng:2020iqw}             & 6491 & $\cdots$ & 6580 & 6607 \\
Ref.~\cite{Gordillo:2020sgc}         & 6351 & $\cdots$ & 6441 & 6471 \\
Ref.~\cite{Yang:2021hrb}             & 6343 & $\cdots$ & 6439 & 6204 \\
Ref.~\cite{Bedolla:2019zwg}          & $\cdots$ & 5883 & 6120 & 6246 \\
Ref.~\cite{Faustov:2021hjs}          & $\cdots$ & 6190 & 6271 & 6367 \\
Ref.~\cite{Zhao:2020nwy}             & 6346 & 6476 & 6441 & 6475 \\
Ref.~\cite{Iwasaki:1975pv}           & $\cdots$ & 6200 & $\cdots$ & $\cdots$ \\
Ref.~\cite{Karliner:2016zzc}         & $\cdots$ & 6192 & $\cdots$ & $\cdots$ \\
Ref.~\cite{Barnea:2006sd}            & $\cdots$ & 6038-6115 & 6101-6176 & 6172-6216 \\
Ref.~\cite{Wang:2018poa}& $\cdots$ & 5990(80) & 6050(80) & 6090(80) \\
Ref.~\cite{Debastiani:2017msn}       & $\cdots$ & 5969 & 6021 & 6115 \\
Ref.~\cite{Berezhnoy:2011xn}         & $\cdots$ & 5966 & 6051 & 6223 \\
Ref.~\cite{Anwar:2017toa}            & $\cdots$ & $<$6140 & $\cdots$ & $\cdots$ \\
\hline
\hline
\end{tabular}
\end{table}

In Fig.~\ref{fig:tetraquark comparison with GEM}, we present the mass spectra of these tetraquark states and compare them with the GEM predictions from Refs.~\cite{Liu:2019zuc,Wang:2026gch}. For the $cc\bar{c}\bar{c}$ system, in Fig.~\ref{fig:tetraquark comparison with GEM}(a), the masses of states $J^P=0^{+}_{\bar{3}3}$, $J^P=1^+$ and $J^P=2^+$ are approximately 20 MeV lower than the results in Ref.~\cite{Liu:2019zuc}.  The mass of the  $J^P=0^{+}_{6\bar{6}}$ state is approximately 101 MeV lower. Due to the enlarged mass splitting induced by configuration mixing, the lower $0^+$ state becomes 133 MeV lower than the results in Ref.~\cite{Liu:2019zuc}, while the higher state becomes 10 MeV higher. For the $bb\bar{b}\bar{b}$ system in Fig.~\ref{fig:tetraquark comparison with GEM}(b), the masses of states $J^P=0^{+}_{\bar{3}3}$, $J^P=1^+$ and $J^P=2^+$ are approximately 35 MeV lower than the results in Ref.~\cite{Liu:2019zuc}.
The $J^P=0^{+}_{6\bar{6}}$ mass is approximately 128 MeV lower. The mixing scheme makes the two $0^+$ masses lower than the results in Ref.~\cite{Liu:2019zuc} by 138 MeV and 27 MeV, respectively. The above results show that for tetraquarks, the expected masses from the NNQS are significantly lower than the GEM predictions in Ref.~\cite{Liu:2019zuc}. This shows that neural-network wave function has a stronger expressive power for the multiquark wave function than the GEM used in Ref.~\cite{Liu:2019zuc}, as indicated by the variational principle. This discrepancy originates from the difference in the expressive power of the multiquark wave functions between the two methods. The GEM relies on predefined Gaussian bases, where the exponentially growing number of required basis in multiquark needs to be truncated. In contrast, the neural networks, serving as a universal approximator of any functions, circumvents the need to predefine specific spatial wave function forms. This capability overcomes the bottleneck of conventional GEM, yielding a better variational upper bound for the ground state energy. 
Specifically, the pure $J^P=0^{+}_{6\bar{6}}$ mass and the corresponding dominated state in NNQS predictions are quite different from the GEM predictions in Ref.~\cite{Liu:2019zuc}. This suggests that the color interactions between the two color-singlet bases $\mathcal{C}_1$ and $\mathcal{C}_2$ in Eq.~\eqref{eq:singlet} can affect the spatial wave function distribution. It suggests that in different color configurations, the Gaussian basis parameters in GEM need to be optimized separately to reflect how the color interactions impact the spatial wave function in each configuration.

It is worth mentioning that, for the $bb\bar{c}\bar{c}$ system in Fig.~\ref{fig:tetraquark comparison with GEM}(c), the masses of the $J^P=0^{+}_{\bar{3}3}$-dominated physical state, $J^P=1^+$ and $J^P=2^+$ are lower than the GEM predictions by Liu et al.~\cite{Liu:2019zuc} in 2019 by approximately 20 to 30 MeV. The mass of the $J^P=0^{+}_{6\bar{6}}$-dominated physical state is lower by approximately 177 MeV. However, when we compare with their GEM predictions in 2026~\cite{Wang:2026gch}, which utilized explicitly correlated Gaussian basis functions, we found that the masses of the $J^P=0^{+}_{\bar{3}3}$-dominated physical state, $J^P=1^+$ and $J^P=2^+$ are very close, i.e. about 4 MeV lower. In addition, the mass of the $J^P=0^{+}_{6\bar{6}}$-dominated state is approximately 27 MeV lower.
This indicates that our proposed method is a more favorable approach.
At the same time, it suggests that the GEM and ECG prediction of the mass of the state $J^P=0^{+}_{6\bar{6}}$ can continue to be improved.

\begin{figure}[htbp]
    \centering
    \includegraphics[width=0.95\linewidth]{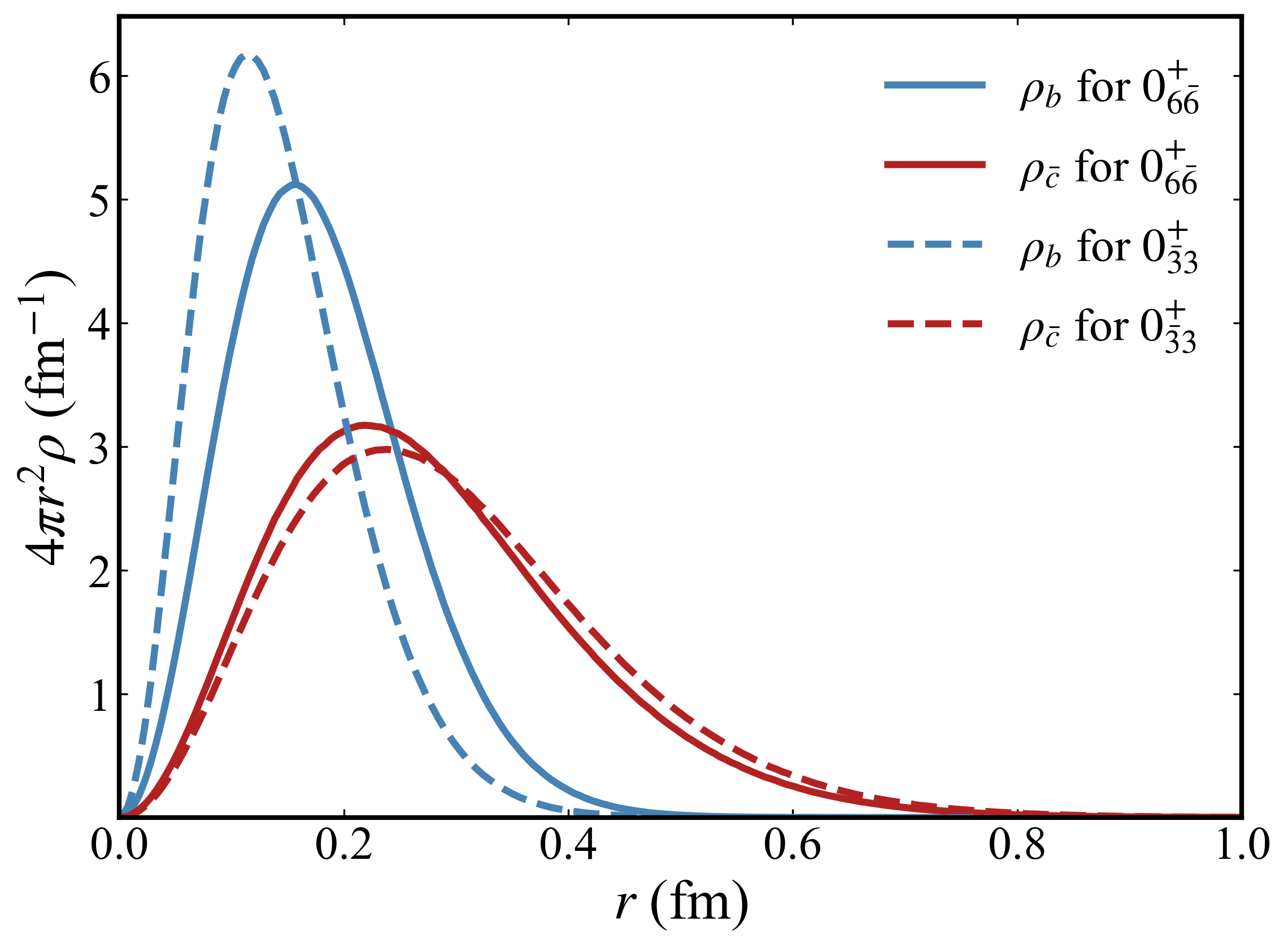}
    \caption{Point particle densities (defined by Eq.~\eqref{eq:rho}) of the $bb\bar{c}\bar{c}$ system for the $0^{+}_{6\bar{6}}$ (solid lines) and $0^{+}_{\bar{3}3}$ (dashed lines) states obtained with NNQS. The results for the bottom quark and anti charm quark are shown in blue and red, respectively.}
    \label{fig:comparison_66_vs_33}
\end{figure}
To show how the color interactions between the two color-singlet bases $\mathcal{C}_1$ and $\mathcal{C}_2$ affect the spatial wave function distribution, we calculate the root-mean-square(RMS) radii $R_{ij}$ for the $0^{+}_{6\bar{6}}$ and $0^{+}_{\bar{3}3}$ states of the $bb\bar{c}\bar{c}$ system, where $R_{ij} = \sqrt{\langle |\bm{r}_i-\bm{r}_j|^2 \rangle}$. 
For the $0^{+}_{6\bar{6}}$ state, the RMS radii are $R_{bb}$ = 0.38 fm, $R_{\bar{c}\bar{c}}$ = 0.49 fm, and $R_{b\bar{c}}$ = 0.39 fm. For the $0^{+}_{\bar{3}3}$ state, the RMS radii are $R_{bb}$ = 0.28 fm, $R_{\bar{c}\bar{c}}$ = 0.46 fm, and $R_{b\bar{c}}$ = 0.41 fm. This indicates that the $0^{+}_{6\bar{6}}$ and $0^{+}_{\bar{3}3}$ states have different spatial distribution structures. Furthermore, we define the point particle density~\cite{Yang:2022rlw},
\begin{equation}
    \rho_{b/\bar{c}}(r) = \frac{1}{4\pi r^2} \frac{\langle \Psi | \sum_{i=1}^2 \delta(|\bar{\bm{r}}_i| - r) | \Psi \rangle}{\langle \Psi | \Psi \rangle},
    \label{eq:rho}
\end{equation}
of the bottom and anti charm quarks. Taking $bb\bar{c}\bar{c}$ system as an example, in Fig.~\ref{fig:comparison_66_vs_33}, we show the results of the calculation, where $\rho_b$ and $\rho_{\bar{c}}$ represent the density of bottom and anti charm quarks, respectively.
In both $0^{+}_{6\bar{6}}$ and $0^{+}_{\bar{3}3}$ states, the distribution of $\rho_{\bar{c}}$ is very close. However, the distribution of $\rho_b$ in $0^{+}_{\bar{3}3}$ state is clearly more compact than its distribution in $0^{+}_{6\bar{6}}$ state. Our calculations show how the color configurations affect the spatial wave function.

\begin{figure*}[!htbp]
    \centering
    \includegraphics[width=\textwidth]{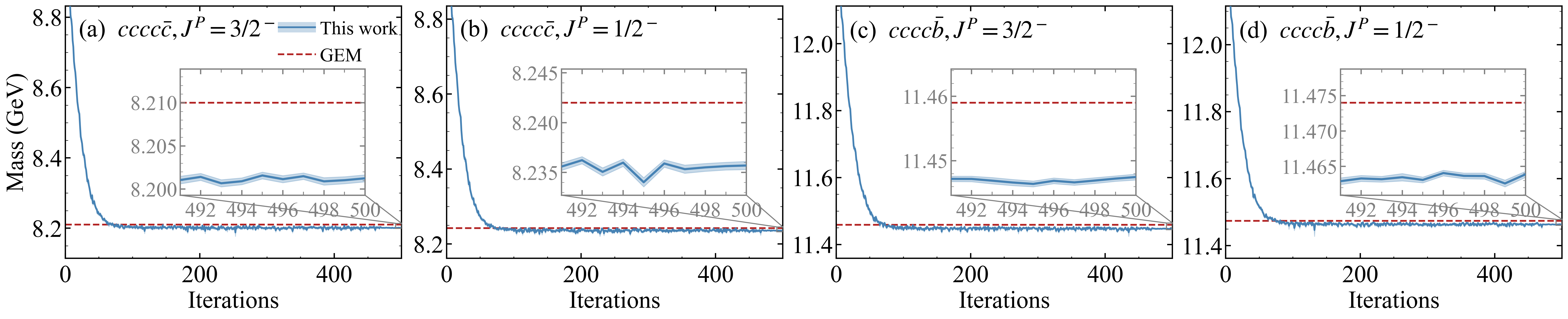}
    \caption{The mass estimate as a function of iteration steps for (a) $J^P=3/2^-$ $cccc\bar{c}$ pentaquark, (b) $J^P=1/2^-$ $cccc\bar{c}$ pentaquark, (c) $J^P=3/2^-$ $cccc\bar{b}$ pentaquark and (d) $J^P=1/2^-$ $cccc\bar{b}$ pentaquark in the optimization progress. The Monte Carlo statistical errors of the masses are shown by the blue shaded area, and the red dashed line represents the GEM prediction by Liang et al.~\cite{Liang:2024met}. The insets in each subplot magnify the final optimization stage.}
    \label{fig:NNpentaquark}
\end{figure*}
\begin{figure*}[!htbp]
    \centering
    \includegraphics[width=\textwidth]{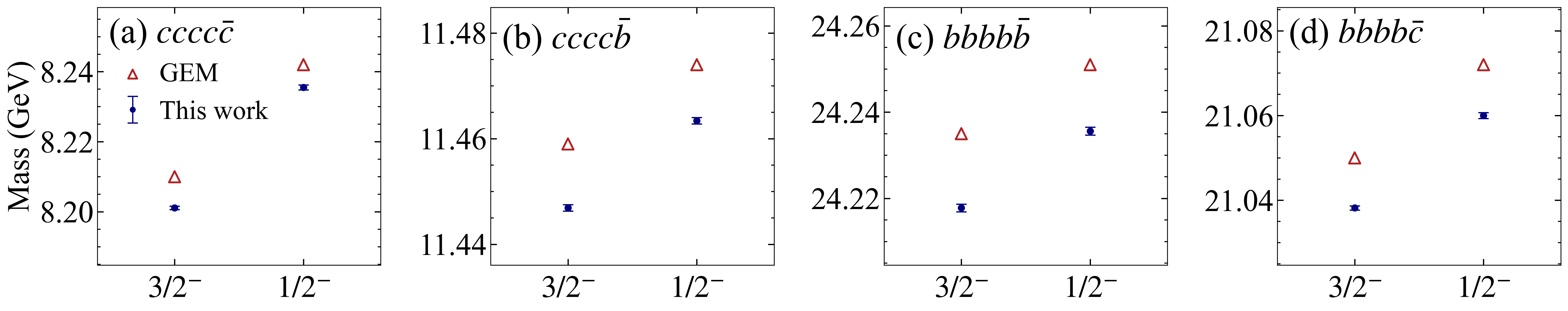}
    \caption{ A comparison of the predicted masses of the pentaquark states $cccc\bar{c}$ (a), $cccc\bar{b}$ (b), $bbbb\bar{b}$ (c), and $bbbb\bar{c}$ (d). The blue points, red triangles represent this work predictions and the GEM predictions by Liang et al.~\cite{Liang:2024met}.
    }
    \label{fig:pentaquark comparison with GEM}
\end{figure*}
Taking the $cc\bar{c}\bar{c}$ system as an example, we compare our results with other model calculations in Table~\ref{tab:tetraquark_masses_compared_with_others} and Fig.~\ref{fig:tetraquark_mass_comparison}. Our predicted masses are slightly lower than those from the nonrelativistic quark models of Refs.~\cite{Liu:2019zuc,Lloyd:2003yc}, which explicitly treat both confining and Coulomb potentials. They are also lower than the MIT bag model result~\cite{Yan:2023lvm} and the diquark model with color‑magnetic interactions~\cite{Wu:2016vtq}. In general, our results agree well with QCD sum rules~\cite{Chen:2016jxd} and several other potential model calculations~\cite{Ader:1981db,Lu:2020cns,Zhang:2022qtp,Wang:2019rdo,Deng:2020iqw,Lloyd:2003yc}. However, our masses are higher than the DMC predictions~\cite{Gordillo:2020sgc} and the lattice‑QCD predictions~\cite{Yang:2021hrb}, and considerably higher than those reported in Refs.~\cite{Bedolla:2019zwg,Deng:2020iqw,Faustov:2021hjs,Zhao:2020nwy,Iwasaki:1975pv,Karliner:2016zzc,Barnea:2006sd,Wang:2018poa,Debastiani:2017msn,Berezhnoy:2011xn,Anwar:2017toa}. These differences arise partly from the use of different potential models~\cite{Deng:2020iqw,Faustov:2021hjs,Zhao:2020nwy} and partly from the omission of explicit confining potentials in some of those studies~\cite{Karliner:2016zzc,Barnea:2006sd,Berezhnoy:2011xn,Anwar:2017toa}.

In this work, we construct antisymmetric spin-color wave functions and symmetric spatial wave functions. This restriction makes our results  above the dimeson threshold. It is worth mentioning that Wu et al.~\cite{Wu:2025wvv} construct the DeepQuark framework to formulate a more general wave function to ensure overall antisymmetry, with their calculated results for fully heavy tetraquark states approaching to the dimeson threshold.
\begin{table}[!htbp]
\setlength{\tabcolsep}{8mm}
\renewcommand{\arraystretch}{2}
\centering
\caption{Predicted mass (MeV) spectra of pentaquarks for $cccc\bar{c}$, $cccc\bar{b}$, $bbbb\bar{b}$ and $bbbb\bar{c}$ systems.}
\label{tab:The masses of pentaquarks}
\begin{tabular}{lccc}   
\hline
\hline
State&$J^{P}$ & Mass \\
\hline
\multirow{2}{*}{$cccc\bar{c}$}   
&$3/2^{-}$      & $8201.1(5)$ \\
&$1/2^{-}$      & $8235.5(7)$ \\
\multirow{2}{*}{$cccc\bar{b}$}  
&$3/2^{-}$      & $11446.9(6)$ \\
&$1/2^{-}$      & $11463.4(6)$ \\
\multirow{2}{*}{$bbbb\bar{b}$}  
&$3/2^{-}$      & $24217.8(9)$ \\
 &$1/2^{-}$     & $24235.6(9)$ \\
\multirow{2}{*}{$bbbb\bar{c}$}   
&$3/2^{-}$      & $21038.2(5)$ \\
&$1/2^{-}$      & $21060.0(7)$ \\
\hline
\hline
\end{tabular}
\end{table}

\begin{figure*}[!htbp]
    \centering
    \includegraphics[width=\textwidth]{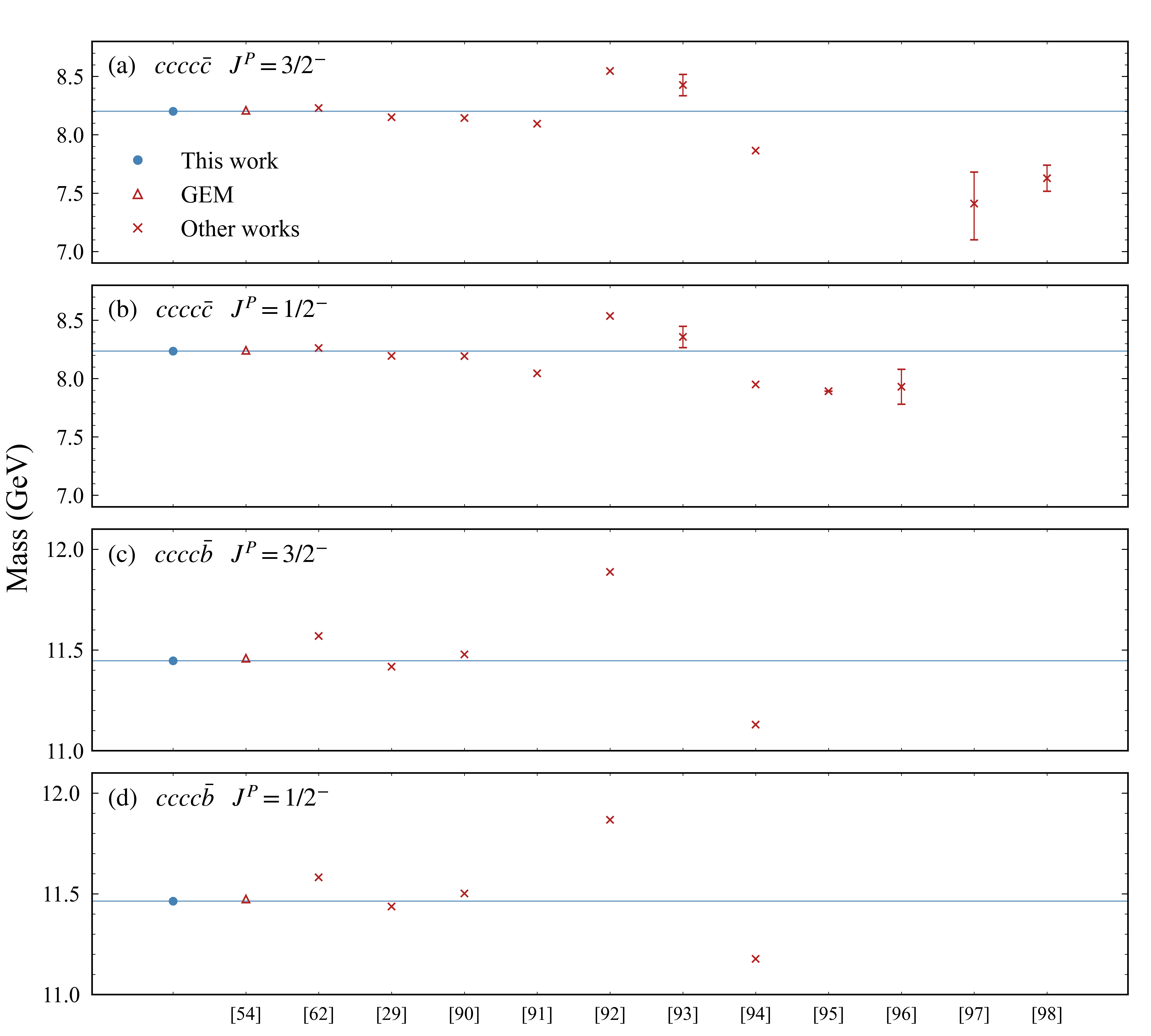}
    \caption{A comparison of the predicted masses of the pentaquark states $cccc\bar{c}$ with $J^P=\frac{3}{2}^-$ (a) and $J^P=\frac{1}{2}^-$ (b), and $cccc\bar{b}$ with $J^P=\frac{3}{2}^-$ (c) and $J^P=\frac{1}{2}^-$ (d). The blue points (horizontal lines), red triangles and red crosses represent the predictions from this work, GEM by Liang et al.~\cite{Liang:2024met}, and other works, respectively.}
    \label{fig:pentaquark_mass_comparison}
\end{figure*} 

Extending our method to fully heavy pentaquarks. The optimization performance of NNQS in pentaquark is shown in Fig.~\ref{fig:NNpentaquark}, and we take the $cccc\bar{c}$ and $cccc\bar{b}$ systems as examples. The mass expectation for the $cccc\bar{c}$ and $cccc\bar{b}$ pentaquarks decreases rapidly during the initial optimization phase, converging stably below the GEM results~\cite{Liang:2024met}. The Monte Carlo statistical uncertainties (blue shaded in Fig.~\ref{fig:NNpentaquark}) are limited and stable, confirming that the NNQS method provides a reliable prediction. All predicted mass spectra for the $cccc\bar{c}$, $cccc\bar{b}$, $bbbb\bar{b}$ and $bbbb\bar{c}$ systems have been given in Table~\ref{tab:The masses of pentaquarks}. We find that the mass of the $J^P=1/2^-$ state is slightly larger than that of the $J^P=3/2^-$ state. In all pentaquark systems, the mass splitting between these two states is about 15 to 35 MeV.

In Fig.~\ref{fig:pentaquark comparison with GEM}, we present the mass spectra of the pentaquark states and compare them with the results from the GEM calculated by Liang et al. in Ref.~\cite{Liang:2024met}. For the $cccc\bar{c}$ system, i.e. Fig.~\ref{fig:pentaquark comparison with GEM}(a), the masses of the $J^P=3/2^-$ and $J^P=1/2^-$ states are approximately 6 to 9 MeV lower than those in Ref.~\cite{Liang:2024met}. For the $cccc\bar{b}$ system, i.e. Fig.~\ref{fig:pentaquark comparison with GEM}(b), the masses of the $J^P=3/2^-$ and $J^P=1/2^-$  states are lower than those in Ref.~\cite{Liang:2024met}  by approximately 10 to 12 MeV. 
For the $bbbb\bar{b}$ system, i.e. Fig.~\ref{fig:pentaquark comparison with GEM}(c), the masses of the $J^P=3/2^-$ and $J^P=1/2^-$ are also lower than those predicted in Ref.~\cite{Liang:2024met} by approximately 15 to 17 MeV. For the $bbbb\bar{c}$ system, i.e. Fig.~\ref{fig:pentaquark comparison with GEM}(d), the masses of the $J^P=3/2^-$ and $J^P=1/2^-$ states are approximately 12 MeV lower than those Ref.~\cite{Liang:2024met}. The above comparisons show that for pentaquarks, the mass expectations from the NNQS remain systematically lower than GEM predictions in Ref.~\cite{Liang:2024met}.
In addition, the above comparison also illustrates that 
the deviations grow with the increasing quark masses, which means that the GEM in Ref.~\cite{Liang:2024met} performs better for charm systems than for bottom systems. 
It once again demonstrates that neural-network wave functions have stronger expressive power than GEM used in Ref.~\cite{Liang:2024met} on multiquark wave functions.

\begin{table}[h]
\setlength{\tabcolsep}{1mm}
\renewcommand{\arraystretch}{1.5}
\centering
\caption{Our predicted masses(MeV) for the $cccc\bar{c}$ and $cccc\bar{b}$ systems compared with others.}
\label{tab:pentaquark masses compared with others}
\begin{tabular}{lcccc}   
\hline
\hline
State & \multicolumn{2}{c}{$cccc\bar{c}$}   & \multicolumn{2}{c}{$cccc\bar{b}$}  \\
$J^P$ & $3/2^{-}$  & $1/2^{-}$  & $3/2^{-}$ & $1/2^{-}$ \\
\hline
This work & 8201.1(5) & 8235.5(7) & 11446.9(6) & 11463.4(6) \\
Ref.~\cite{Liang:2024met} & 8210 & 8242 & 11459 & 11474 \\
Ref.~\cite{Zhang:2023hmg} & 8229 & 8262 & 11569 & 11582 \\
Ref.~\cite{Gordillo:2024blx} & 8151 & 8194 & 11417 & 11437 \\
Ref.~\cite{An:2022fvs} & 8144.6 & 8193.2 & 11477.8 & 11501.5 \\
Ref.~\cite{Yang:2022bfu} & 8095 & 8045 & $\cdots$ & $\cdots$ \\
Ref.~\cite{Rashmi:2024ako} & 8547.4 & 8537.4 & 11887.6 & 11867.7 \\
Ref.~\cite{Sharma:2025grd} &  8425.9(911) & 8356.9(910) & $\cdots$ & $\cdots$ \\
Ref.~\cite{An:2020jix} & 7864 & 7949 & 11130 & 11177 \\
Ref.~\cite{Yan:2021glh} & $\cdots$ & 7892.3(4) & $\cdots$ & $\cdots$ \\
Ref.~\cite{Wang:2021xao} & $\cdots$ & 7930(150) & $\cdots$ & $\cdots$ \\
Ref.~\cite{Zhang:2020vpz} & $7410^{+270}_{-310}$ & $\cdots$ & $\cdots$ & $\cdots$ \\
Ref.~\cite{Azizi:2024ito} & 7628(112) & $\cdots$ & $\cdots$ & $\cdots$ \\
\hline
\hline
\end{tabular}
\end{table}

Taking the $cccc\bar{c}$ and $cccc\bar{b}$ pentaquarks as examples, we compare our results with other model calculations in Table~\ref{tab:pentaquark masses compared with others} and Fig.~\ref{fig:pentaquark_mass_comparison}. Our predicted masses are consistent with several recent calculations~\cite{Liang:2024met,Zhang:2023hmg,An:2022fvs,Gordillo:2024blx,Yang:2022bfu}. Specifically, they are slightly lower than the GEM results~\cite{Liang:2024met} and the MIT bag model~\cite{Zhang:2023hmg}, but slightly higher than those from the single Gaussian variational method~\cite{An:2022fvs}, the DMC method~\cite{Gordillo:2024blx}, and the GEM combined with a complex-scaling approach~\cite{Yang:2022bfu}. In addition, our results are lower than the calculations based on the screened charge scheme~\cite{Rashmi:2024ako} and the extended Gürsey-Radicati formalism~\cite{Sharma:2025grd}. On the other hand, our masses are significantly higher than those evaluated via the chromomagnetic interaction (CMI) model~\cite{An:2020jix}, the chiral quark model~\cite{Yan:2021glh}, and QCD sum rules~\cite{Wang:2021xao,Zhang:2020vpz,Azizi:2024ito}. These discrepancies largely originate from the different choices of the potentials employed in these models. Importantly, if the same potential were adopted, our variational approach would produce results closer to the true values, demonstrating its advantage in providing accurate predictions.

\section{Summary}
In this work, we propose a computational framework based on neural-network quantum states to calculate the mass spectra of fully heavy multiquark systems within a nonrelativistic quark model. The color‑spin wave functions are constructed exactly via $\mathrm{SU}(3)\times \mathrm{SU}(2)$ group theory to enforce fermionic antisymmetry, while the spatial part is parametrized by deep neural networks. After benchmarking on conventional mesons and baryons—where NNQS reproduces or improves upon the Gaussian expansion method results—we compute the $S$-wave ground‑state masses of fully heavy tetraquarks and pentaquarks. Our approach consistently yields lower energies than GEM, a direct consequence of the variational principle, indicating that the neural network wave function offers greater expressive power by adaptively capturing nonperturbative many‑body spatial correlations without relying on restrictive a priori assumptions. Compared with traditional methods, NNQS overcomes the exponential growth of basis size and avoids the fermion sign problem, making it particularly suitable for high‑dimensional strongly correlated systems. Our results show good agreement with various model calculations, and importantly, when the same  potential is employed, NNQS produces values closer to the true ones, demonstrating its superior accuracy and reliability. This framework not only provides a robust tool for the spectroscopy of exotic hadrons but also offers a flexible continuous‑space solver for investigating microscopic binding mechanisms. Future extensions to hexaquark and more complex systems are anticipated, which will further aid experimental searches for all‑heavy multiquark states.

\section{Acknowledgments}
We are grateful to Zi-Xiao Zhang, Hong-Fei Zhang and Jifeng Hu for code support, and to Xin-Yue Hu, Zhan-Wei Liu, Xiao-Hai Liu and Lu Meng for valuable discussions on multiquark. This work is partly supported by the National Natural Science Foundation of China with Grants Nos.~12375073, 12547105, and 12342502. 

%Bibliography
\nocite{*}
%\bibliography{ref.bib}  

\onecolumngrid
\newpage
\appendix

\clearpage
\section{The hyperparameters used in our calculations}\label{sup:neural-network}
In the Appendix, we explain and summarize the architecture of the neural network and the hyperparameters used in our calculations. We employ a fully connected neural network. Specifically, it is a fully connected neural network composed of an input layer, multiple hidden layers, and an output layer. The number of nodes corresponds to the number of neurons. The output of the previous layer serves directly as the input to the subsequent layer. Specifically, each hidden layer applies a nonlinear activation function:
\begin{equation}
    \bar{\bm{r}}^{(l+1)} = f(\bm{\omega}^{(l)}\bar{\bm{r}}^{(l)}+b^{(l)}),
\end{equation}
where $\bm{\omega}^{(l)}$ and $b^{(l)}$ denote the weights and biases of the $l$th layer, respectively, and $f$ is the activation function. We employ the SiLU as the activation function in this work. Its specific form is expressed as:
\begin{equation}
    \text{SiLU}(x) = \frac{x}{1 + e^{-x}}.
\end{equation}
The number of hidden layer nodes and the number of layers control the expressive power of the neural network wave function. A higher number of nodes and layers yields a more expressive neural network, but incurs a higher computational cost. For meson and  baryon systems, the network consists of 4 hidden layers with 24 nodes per layer. For tetraquark and pentaquark systems, the network has 6 hidden layers with 32 nodes each. This design ensures sufficient expressive power while preserving computational feasibility.
The hyperparameters for our calculations are summarized in Table~\ref{tab:hyperparameters}.
\begin{table}[htbp]
\setlength{\tabcolsep}{10mm}
\renewcommand{\arraystretch}{1.5}
\centering
\caption{The hyperparameters used in our calculations.}
\label{tab:hyperparameters}
\begin{tabular}{lc lc}   
\hline
\hline
Parameter & Value & Parameter & Value \\
\hline
Acceptance rate range & 0.75--0.85 & Regularization constant & $10^{-3}$--$10^{-4}$ \\
Acceptance rate factor & 1.15 & Batch size & 2000 \\
Steps & 800 & Large batch size & 80000 \\
Initial learning rate & 0.04--0.05 & Large batch iterations & 30 \\
Total iterations & 500 & Final average iterations & 10 \\
\hline
\hline
\end{tabular}
\end{table}

In the Metropolis-Hastings Monte Carlo sampling, a new position $r'$ is randomly proposed and then accepted or rejected based on acceptance rate calculated by:
\begin{align}
    & P(r)=\frac{|\Psi(r)|^2}{\int dr|\Psi(r)|^2}, \\
    & P_{\text{accept}}(r\to r') = \text{min} \left \{ 1, \frac{P(r)}{P(r')} \right \}. 
\end{align}
To ensure an efficient and thorough exploration of the space, the sampling step size is adjusted. If the acceptance rate falls outside the acceptance rate range, the step size is multiplied or divided by the acceptance rate factor. The steps parameter refers to the number of random walk steps performed per neural network iteration. The total iterations parameter is the total number of training iterations for the neural network, and the regularization constant refers to the $\epsilon$ in Eq.~\eqref{Parameter Update Rule}. The batch size in our work is the number of Metropolis-Hastings samples. To reduce statistical uncertainty, during the last large batch iterations, we increase the sample size from batch size to the large batch size. Finally, we average the results over the final average iterations to obtain our final results.

\end{document}